\documentclass[a4paper,11pt]{article}
\usepackage{latexsym} 
\usepackage{amsfonts, amsthm, amsmath,amssymb}
\usepackage[a4paper, total={6.3in, 10.5in}]{geometry}
\def\CC{{\mathbb C}}
\def\XX{{\mathbb X}}
\newtheorem{proposition}{Proposition}[section]
\usepackage{times}
\usepackage[T1]{fontenc} 

\title{\sf\bfseries Spectral interaction between universes}

\author{\sf\bfseries A. Bochniak, A. Sitarz \\[3mm]
\sffamily Institute of Theoretical Physics, Jagiellonian University,\\
\sffamily prof.\ Stanis\l awa \L ojasiewicza 11, 30-348 Krak\'ow, Poland \\[4mm]
\sffamily \small E-mail: arkadiusz.bochniak@doctoral.uj.edu.pl, andrzej.sitarz@uj.edu.pl}

\begin{document}
	\maketitle
	
	\abstract{We derive a perturbative formula for the direct interaction between two four-dimensional geometries. Based on the spectral action principle we give an explicit potential up to the third order perturbation around the flat vacua. We 
		present the leading terms of the interaction as polynomials of the invariants of the two metrics and compare the expansion 
		to the models of bimetric gravity.}
\section{Introduction}
One of the most significant achievements of modern physics is geometry's spectacular success in describing the large-scale structure and the evolution of the Universe thanks to general relativity. On the subatomic scale, the geometric picture of gauge theories establishes the natural framework for fundamental particle interactions. Although the common unifying scheme for both, apparently different, types of interactions is not yet known, there exist various approaches that aim to bridge the gap. Noncommutative geometry, which changes the way of approach by making the differential operators as fundamental objects can, at least on the classical level, treat the gauge fields as well as the metric as different fields that parametrize the real physical object, the Dirac operator \cite{Co94,Co96,CoMa07}.

The theory, when applied to the Standard Model of particle interactions can explain, in a purely geometric way, the existence of the Higgs field and the appearance of symmetry-breaking potential. However, the necessary element, that has to be added, includes a geometry of discrete-type, which is described as a finite-dimensional matrix algebra $\mathbb{C} \oplus \mathbb{H} \oplus M_3(\mathbb{C})$. In a simplified model (which ignores the strong interactions and treats the weak interactions as electromagnetic) one can reduce this algebra to $\mathbb{C} \oplus \mathbb{C}$ leading to a simple geometry of a product of the four-dimensional spin manifold with two points \cite{KaSch97,CoLo91}. 

The model, which looks like a two-sheeted geometry and can be compared to the model with two four-dimensional branes or the boundary of a thin domain wall of five dimensions in the bulk (see \cite{DzFM10} for vast literature on the latter topic).  This extends the image of the universe as a brane in the bulk with the possibility of the system of a pair of interacting branes. Since the interactions with the bulk and between the branes influence the physics it is natural to ask what is the origin of the interactions and whether it is possible to have it of purely geometric origin. The answer comes
again through the tools noncommutative geometry. We assume that the interaction between the fields on the two sheets is mediated by the Higgs field, which itself is related to the metric and the connection on the discrete component of the geometry. Following this idea, we can, in principle, derive an explicit and unambiguous interaction between the geometries alone, depending only on the metrics.

In the constructions so far, one usually assumed the natural product-type geometry (product-type Dirac operators), which, after applying the spectral action procedure \cite{EckIo18,Vas03} led to the standard Einstein-Hilbert action for the metric, identical on the two universes. 
Yet this is not the most general form of the Dirac operator and different metrics on the two separate universes are admissible \cite{Si19}. 
Together with the Higgs-type field that mediates between the two geometries one can obtain an interaction term between the two metrics, leading to an interesting class of models, which appear to be viable from the point of view of cosmological models \cite{BoSi21}. 

The general type of the interaction term for two arbitrary metrics is not explicitly computable even in the simple setup. An exact 
answer was obtained only for the Friedmann-Lema{\^i}tre-Robertson-Walker (FRLW) type of Euclidean metrics \cite{BoSi21} which allowed to study the 
stability of solutions of cosmological evolution equations. Interestingly, the obtained models resemble the so-called bimetric gravity \cite{HaRo01}, which is a good candidate to potentially solve the puzzle of dark matter in accordance with the cosmological data \cite{AKMS13,StSc16,SSEM12} and
does not suffer from Boulware-Deser ghost problem \cite{HaRo01,HaRo02}.  However, the bimetric gravity lacks a geometrical interpretation
and the second metric-like field is not well justified from the standpoint of Riemannian geometry. 

As the noncommutative geometry motivated model of two-sheeted space with two metrics yields a similar theory, with a full diffeomorphism invariance, it is natural to ask how do these models differ. In particular, in contrast to bimetric theory, the two-sheeted geometry spectral action 
principle fixes uniquely (up to multiplicative constant) the interaction term between the metrics.  Our previous analysis of the FLRW type 
geometries allowed us to give only a partial answer about the similarities and differences of the two approaches. 

In this note, we derive an explicit form of the spectral action for the infinitesimal perturbation of the flat metric (in the Euclidean setup) on the two-sheeted geometry (up to the fourth-order) and compare it with the general action proposed for bimetric gravity up to the first three orders. This demonstrates that a simple model of noncommutative geometry allows direct and generic interaction between universes (branes) and opens a possibility to study the general properties of such models.

\section{The interaction of geometries - a general construction}
\label{sec:DG}
Noncommutative geometry allows us to generalize classical concepts from differential geometry in a systematic way.  The fundamental 
object is a  spectral triple, which is a system $(A,H,D)$ consisting of a unital $\ast$-algebra represented (faithfully) on a Hilbert 
space $H$ and the Dirac operator $D$, which is essentially self-adjoint on $H$. 
Several compatibility conditions are assumed, such as the compactness of the Dirac operator's resolvent, the boundedness 
of certain commutators, and so on. The spectral triple $(C^{\infty}(\mathcal{M}), L^2(\mathcal{M},g), D)$ encoding the geometry of the 
(compact, spin) Riemannian manifold $(\mathcal{M},g)$ is the canonical example. Locally, $D=i\gamma^\mu(\partial_\mu+\omega_\mu)$, 
where $\omega_\mu$ is the spin connection on the spinor bundle over $(\mathcal{M},g)$. Another example is almost-commutative geometry, which is defined as the product of the canonical spectral triple and some finite one, $(A_F,H_F,D_F)$, with $A_F$ and $H_F$ being finite dimensional, and $D_F$ being a (matrix) operator acting on $H_F$. The corresponding Dirac operator has the form $D\otimes 1+\gamma_M\otimes D_F$ with $\gamma_M$ being the canonical grading on $\mathcal{M}$ (that is, at a given point, $\gamma_M=\gamma_5$ in the associated Clifford algebra). Spectral triples of this product type of geometry were successfully applied to the description of the Standard Model of particle physics \cite{WvS15} and provided a geometric understanding of this model. 

The natural generalization of the almost-commutative product-like geometry for the Riemannian (four-dimensional) manifold $\mathcal{M}$ and the finite space $\mathbb{Z}_2$ is the one with the Dirac operator not being of the product type. This defines 
 the so-called {\it doubled geometry} \cite{Si19, BoSi21}. More precisely, for this spectral triple the Dirac operator is taken 
 to be of the form 
\begin{equation}
\mathcal{D}=\begin{pmatrix}
D_1 & \gamma \Phi\\
\gamma \Phi^\ast& D_2
\end{pmatrix},
\end{equation}
where $D_1$ and $D_2$ are the two usual Dirac operators for the two copies of the (spin) Riemannian manifold $\mathcal{M}$, with metrics $g_1$ and $g_2$, respectively. Here $\gamma$ is an operator that squares to $\kappa=\pm 1$ and is a generalization of the usual grading on the canonical spectral triple (see \cite{BoSi21} for detailed discussion of the role and origin of this operator). We assume that $\gamma$ is Hermitian and anticommutes with all the $\gamma^a$ matrices which are taken to be anti-Hermitian and satisfy $\gamma^a\gamma^b+\gamma^b\gamma^a=-2\delta^{ab}$.

For a given metric $g$ on the manifold $\mathcal{M}$, the Dirac operator can be written explicitely as
\begin{equation}
D=\gamma^a dx^\mu(\theta^a)\frac{\partial}{\partial x^\mu} +\frac{1}{4} \gamma^c \omega_{cab}\gamma^a\gamma^b , 
\end{equation}
where $\{\theta^a\}$ is the orthogonal coframe, $ds^2=g_{\mu\nu}dx^\mu dx^\nu=\theta^a\theta^a$, and the (coefficients of the) spin connection can be computed by using the relation $d\theta^a=\omega^{ab}\wedge \theta^b$.

\section{The perturbative interaction of two metric geometries.}
We assume that our manifold is a four-dimensional Euclidean torus \footnote{As all of the obtained terms are local, this assumption is only technical and the results will hold for any compact manifold.}  with the natural choice of global coordinates and with the metric that is an infinitesimal perturbation of the flat one, 
\begin{equation}
g_{ij}=\delta_{ij} +\epsilon h_{ij},
\end{equation}
with some $h_{ij}$ and the perturbation parameter $\epsilon$.  The inverse metric, $g^{ij}$, is, up to $\epsilon^4$:
\begin{equation}
	g^{ij}=\delta^{ij} -\epsilon h^{ij} + \epsilon^2 h^{ik} h_{k}^{\ j} - \epsilon^3  h^{ik} h_{km} h^{mj}+\epsilon^4 h^{i}_{\ j}h^{jl}h_{lm}h^{mk}.
\end{equation}
In what follows we will be interested in the form of the potential term, therefore we are allowed to put $h_{ij}=\mathrm{const}$ as 
no derivatives of the metric enter. Therefore, the Dirac operator $D$, again up to $\epsilon^4$, becomes:
\begin{equation}
D = \gamma^i \biggl(\delta^{j}_{\ i} - \frac{1}{2} \epsilon h^{j}_{\ i} 
+ \frac{3}{8} \epsilon^2 h^{j}_{\ k} h^{k}_{\ i} 
- \frac{5}{16} \epsilon^3 h^j_{\ k} h^k_{\ l} h^l_{\ i}   + \frac{35}{128}\epsilon^4 h^{j}_{\ l} h^{l}_{\ k}h^{k}_{\ n}h^{n}_{\ i}\biggr) \partial_j.
\end{equation} 

Since we are working with the constant metric we take as the Hilbert space the two copies of the square-summable sections of the usual spinor bundle over the four-torus with respect to the flat metric, ${\mathcal H} = L^2(S) \otimes \CC^2$. This facilitates the computations and does not single out any of the geometries as a preferred one. 

For the two-sheeted metric geometry with the Higgs-type field, as described in Sec. \ref{sec:DG}, the Dirac-type operator in noncommutative geometry has the form,
\begin{equation}
\begin{aligned}	
	\mathcal{D}\; =\; & \gamma^j \partial_j +\gamma F 
	-\frac{1}{2}  \epsilon \gamma^i H^j_{\ i}\partial_j
	+\frac{3}{8}  \epsilon^2 \gamma^i H^{jk} H_{ki} \partial_j \\
	&\; - \frac{5}{16} \epsilon^3 \gamma^i H^j_{\ k} H^k_{\ l} H^l_{\ i} \partial_j + \frac{35}{128}\epsilon^4 
	   \gamma^i H^j_{\ k} H^k_{\ l}H^{l}_{\ n}H^{n}_{\ i}\partial_j,
\end{aligned}	
\end{equation}
where
\begin{equation}
	H_{jk}=\begin{pmatrix}
		h_{1jk} &\\
		& h_{2jk}
	\end{pmatrix}, \qquad F=\begin{pmatrix}
		&\Phi \\
		\Phi^\ast &
	\end{pmatrix},
\end{equation}
and we can also assume that the field $\Phi$ is constant (since we do not investigate the dynamical terms). As a result,
\begin{equation}
\begin{aligned}
	\mathcal{D}^2=& -\partial_j^2 	+\kappa F^2
	+ \epsilon H^{jk} \partial_{j}\partial_k 
    - \frac{\epsilon}{2}\gamma \gamma^j [F, H^k_{\ j}]\partial_k \\
	& -  \epsilon^2 H^{j}_{\ l} H^{lk} \partial_{j} \partial_k  
	   + \frac{3}{8} \epsilon^2\gamma \gamma^j [F, H^{kl} H_{lj}] \partial_k \\
	 & +   \epsilon^3 H^{j}_{\ n} H^{n}_{\ l} H^{lk} \partial_{j} \partial_k  
	     - \frac{5}{16}  \epsilon^3 \gamma \gamma^j [F, H^{k}_{\ l} H^l_{\ n} H^n_{\ j}]\partial_k\\
	     &- \epsilon^4 H^j_{\ m}H^{m}_{\ l}H^l_{\ n}H^{nk}\partial_j\partial_k +\frac{35}{128}\epsilon^4 \gamma\gamma^j[F, H^{k}_{\ l}H^l_{\ m}H^{m}_{\ n}H^n_{\ j}]\partial_k, \\
\end{aligned}	
\end{equation}
where $\kappa=\pm 1$ depending on the properties of the grading $\gamma$.

The computation of the interaction term follows the general principle of the spectral action. Technically, to obtain the 
Einstein-Hilbert action and its generalization we compute the Wodzicki residue of the inverse of $\mathcal{D}^2$ \cite{gilkey}.
The procedure uses the explicit computation of the symbols of the pseudodifferential operator $\mathcal{D}^{-2}$
and the integration over the cosphere.

The homogeneous parts of the symbol of the differential operator $\mathcal{D}^2$ are,
\begin{equation}
	\begin{aligned}
		\mathfrak{a}_2=&\|\xi\|^2 -\epsilon H^{jk}\xi_j\xi_k 
		+ \epsilon^2 H^j_{\ l} H^{lk} \xi_{j} \xi_k
		 -  \epsilon^3 H^j_{\ n} H^n_{\ l} H^{lk} \xi_{j} \xi_k  +\epsilon^4 H^j_{\ m}H^m_{\ l}H^l_{\ n}H^{nk}\xi_j\xi_k, \\
		\mathfrak{a}_1=&-\frac{i}{2}\epsilon \gamma\gamma^j [F,H^{k}_{\ j}]\xi_k
		+ \frac{3i}{8} \epsilon^2\gamma \gamma^j [F, H^{kl} H_{lj}] \xi_k
		- \frac{5i}{16}  \epsilon^3 \gamma \gamma^j [F, H^{k}_{\ l} H^l_{\ n} H^n_{\ j}]\xi_k\\
		&+\frac{35i}{128}\epsilon^4\gamma\gamma^j [F,H^k_{\ l}H^l_{\ m}H^m_{\ n}H^n_{\ j}]\xi_k, \\
		\mathfrak{a}_0=&\kappa F^2.
	\end{aligned}
\end{equation}

The symbols of its inverse, $\sigma_{\mathcal{D}^{-2}}=\mathfrak{b}_{-2}+\mathfrak{b}_{-3}+ \mathfrak{b}_{-4}+ \ldots$, are
much more complicated, with the principal symbol,
\begin{equation}
\begin{aligned}&
\mathfrak{b}_{-2} = \frac{1}{\|\xi\|^2}\left(1 + \epsilon H^{jk}\frac{\xi_j\xi_k}{\|\xi\|^2}
+ \epsilon^2\left(H^{jk}H^{mn} \frac{\xi_j\xi_k\xi_m\xi_n}{\|\xi\|^4} -H^j_{\ l}H^{lk}\frac{\xi_j\xi_k}{\|\xi\|^2} \right) \right. \\
\quad &\left.  + \, \epsilon^3 \left(
H^{jk}H^{mn} H^{rs}  \frac{\xi_j\xi_k\xi_m\xi_n \xi_r\xi_s }{\|\xi\|^6}
- 2  H^{jk}H^{m}_{\ l} H^{ln}  \frac{\xi_j \xi_k  \xi_m\xi_n }{\|\xi\|^4}
+ H^j_{\ n} H^n_{\ l} H^{lk} \frac{\xi_j \xi_k }{\|\xi\|^2} \right)\right. \\
\quad&\left. +\epsilon^4\left( H^{jk}H^{mn}H^{rs}H^{pq}\frac{\xi_j\xi_k\xi_m\xi_n\xi_r\xi_s\xi_p\xi_q}{\|\xi\|^8}- 3H^j_{\ l}H^{lk}H^{mn}H^{rs}\frac{\xi_j\xi_k\xi_m\xi_n\xi_r\xi_s}{\|\xi\|^6}\right.  \right. \\
\quad&\left.\left. + \left(2H^{j}_{\ n}H^{n}_{\ l}H^{lk}H^{rs}+H^j_{\ l}H^{lk}H^r_{\ n}H^{ns}\right) \frac{\xi_j\xi_k\xi_r\xi_s}{\|\xi\|^4} - H^j_{\ m }H^m_{\ l} H^{l}_{\ n}H^{nk}\frac{\xi_j\xi_k}{\|\xi\|^2}\right)  \right).
\end{aligned}	
\end{equation}

For the $\mathfrak{b}_{-4}$ we are, effectively, interested only in its component $\mathfrak{b}_{-4}'$ that contains the interaction terms between the metrics (there will be terms that are proportional to the volume and the separate Einstein-Hilbert terms for each metric) that arises (for the constant metrics) exclusively from the product $ \mathfrak{b}_{-2} \mathfrak{a}_1 \mathfrak{b}_{-2} \mathfrak{a}_1 \mathfrak{b}_{-2}$ term. To obtain the final expression we already use the properties of the trace over the algebra of $2\times 2$ matrices as well as over the Clifford algebra, which significantly simplifies the number of terms. 

For this part, we  obtain,
\begin{equation}
\begin{aligned}
& \mathrm{Tr}_{Cl} \mathrm{Tr} (\mathfrak{b}_{-4}') = -\frac{\kappa}{\|\xi\|^6}\biggl\{\epsilon^2\, 
\mathrm{Tr}\left([F,H^m_{\ j}][F,H^{nj}]\right)\xi_m\xi_n\\
+ &\epsilon^3 \,\mathrm{Tr}\left(3[F,H^k_{\ j}] [F,H^{nj}] H^{st}    
      \frac{\xi_k\xi_n\xi_s\xi_t}{\|\xi\|^2}
   -\frac{3}{2}[F,H^k_{\ j}][F,H^n_{\ m} H^{mj} ] \xi_n\xi_k\right) \\
+& \epsilon^4\, \mathrm{Tr} 
\left[  \left( 4 [ F,H^k_{\ j}] [F,H^{pj}] H^{rs}H^{mn}
  + 2 [ F,H^k{\ j}] H^{rs} [F,H^{pj}] H^{mn} \right) 
\frac{\xi_k\xi_p\xi_r\xi_s\xi_m\xi_n}{\|\xi\|^4}  \right. \\
& - 3 [F,H^k{\ j} ][F,H^{pj}]  H^{r}_{\ n} H^{ns} 
\frac{\xi_k\xi_p\xi_r\xi_s}{\|\xi\|^2} \\
&- \frac{9}{4}
\left( [F,H^k_{\ j}] [F,H^{nm} H_{m}^{\ j}] H^{rs}
       + [F,H^{nm} H_{m}^{\ j}]  [F,H^k_{\ j}] H^{rs} \right)
\frac{\xi_k\xi_n\xi_r\xi_s}{\|\xi\|^2} \\
&+  \frac{5}{4}[F,H^k_{\ j}][F,H^{n}_{\ m} H^{m}_{\ r} H^{rj}] \xi_k\xi_n  + \frac{9}{16}[F,H^{kl}H_{lj}][F,H^{np}H_{pj}]
      \xi_k\xi_n  \biggr]\biggr\}.
\end{aligned}
\end{equation}
After integrating the result over the cosphere (the integrals we have used are in the appendix \ref{app:A}) and the manifold, and further using the symmetry of the perturbation terms, the interaction term reduces to,
\begin{equation}
\label{Sh1h2}
\begin{aligned}
 S(h_1,h_2) \sim &  \epsilon^2 \, \mathrm{Tr} (h_2-h_1)^2 \\
 + &\epsilon^3 \left[\frac{1}{4}\mathrm{Tr}(h_2-h_1)^2\mathrm{Tr}(h_1+h_2)
 -  
  \mathrm{Tr}\left[(h_2-h_1)^2(h_1+h_2)\right]\right] \\
+&\epsilon ^4 \left\{\frac{1}{24}\mathrm{Tr}(h_2-h_1)^2 \left[(\mathrm{Tr}\, h_1)^2 + (\mathrm{Tr}\, h_2)^2
+(\mathrm{Tr}\, h_1)(\mathrm{Tr}\, h_2) \right. \right.\\
& \left. \qquad \qquad -4\mathrm{Tr}(h_1^2+h_2^2)+2\mathrm{Tr}(h_1h_2)\right]\\
&\left. - \frac{1}{6}\mathrm{Tr}(h_2-h_1)^4+\frac{5}{4}\mathrm{Tr}\left[(h_2-h_1)(h_2^3-h_1^3)\right]-\frac{3}{16}\mathrm{Tr}(h_2^2-h_1^2)^2\right.\\
&\left.+\frac{1}{12}\left[ (\mathrm{Tr}\, h_1)\mathrm{Tr}[(h_2-h_1)^2 h_1]+(\mathrm{Tr}\, h_2)\mathrm{Tr}[(h_2-h_1)^2h_2]\right]\right.\\
&\left.-\frac{7}{24}\mathrm{Tr}(h_1+h_2)\mathrm{Tr}[(h_2-h_1)^2(h_2+h_1)]
\right\}.
\end{aligned}
\end{equation}
This expression has a much simpler form when replacing the $h_1,h_2$ perturbations by their linear
combinations. With 
$$W_- =  h_2 - h_1, \qquad \qquad W_+ = h_2 + h_1,$$
we have:
\begin{equation}
\label{eq:bimW}
\begin{aligned}
S(h_1,h_2) &\sim \epsilon^2  \, \mathrm{Tr} (W_-)^2 \\
               +&\epsilon^3  \,  	\frac{1}{4}\left( 
      	     \mathrm{Tr}(W_-)^2 \mathrm{Tr}(W_+)
		- 4 \mathrm{Tr} (W_+ W_-^2 ) \right) \\
		+&\epsilon^4 \, \left(\frac{1}{32}\mathrm{Tr}(W_-^2)\left(\mathrm{Tr}\, W_+\right)^2+\frac{1}{96}\mathrm{Tr}(W_-^2)\left(\mathrm{Tr}\, W_-\right)^2-\frac{1}{16}\mathrm{Tr}(W_-^2)\mathrm{Tr}(W_+^2)\right.\\
		&\left. -\frac{5}{48}\left(\mathrm{Tr}\, W_-^2\right)^2+\frac{7}{48}\mathrm{Tr}(W_-^4)+\frac{3}{4}\mathrm{Tr}(W_+^2W_-^2)\right.\\
		&\left.+\frac{1}{24}\mathrm{Tr}(W_-)\mathrm{Tr}(W_-^3)
		   -\frac{1}{4}\mathrm{Tr}(W_+)\mathrm{Tr}(W_-^2W_+)\right).
\end{aligned}
\end{equation}

\section{Comparison with bimetric gravity models}
The commonly assumed interaction part between the two metrics $g_1$, $g_2$ in the bimetric gravity models \cite{HaRo01,AKMS13} is of the form
\begin{equation}
S_{\mathrm{int}}\sim\int d^4x \sqrt{\det g_2}\left(\sum\limits_{n=0}^4\beta_n e_n(\mathbb{X}) \right),
\end{equation}
where the matrix $\mathbb{X}=\sqrt{g_2^{-1}g_1}$, and the constants  $\beta_n$ are free parameters of the model.
The invariant functions $e_n$ are,
\begin{subequations} 
$$	
\begin{aligned}
e_0(\mathbb{X})&=1,  
& e_1(\mathbb{X})&=\mathrm{Tr}(\mathbb{X}),\\
e_2(\mathbb{X})&=\frac{1}{2}\left((\mathrm{Tr}(\mathbb{X}))^2-\mathrm{Tr}(\mathbb{X}^2)\right), \qquad \qquad
& e_4(\mathbb{X})&=\det(\mathbb{X}), \\
\end{aligned}
$$ 
$$
\begin{aligned}
e_3(\mathbb{X})&=\frac{1}{6}\left(\left(\mathrm{Tr}(\mathbb{X})\right)^3-3\mathrm{Tr}(\mathbb{X})\mathrm{Tr}(\mathbb{X}^2)+2\mathrm{Tr}(\mathbb{X}^3)\right).
\end{aligned}
$$
\end{subequations}
Let us expand the above action in $\epsilon$ when $g_{1ij}=\delta_{ij}+\epsilon h_{1ij}$ and $g_{2ij}=\delta_{ij}+\epsilon h_{2ij}$.
First, we compute,
\begin{equation}
\begin{aligned}
\sqrt{\det g_2} &= 1+\frac{1}{2}\epsilon\mathrm{Tr}(h_2)
+\frac{1}{8}\epsilon^2\left((\mathrm{Tr}(h_2))^2-2\mathrm{Tr}(h_2^2)\right) \\
& \quad +\frac{1}{48}\epsilon^3\left[(\mathrm{Tr}(h_2))^3-6\mathrm{Tr}(h_2)\mathrm{Tr}(h_2^2)
  +8\mathrm{Tr}(h_2^3)\right],	
\end{aligned}	
\end{equation}
and 
\begin{equation}
g_2^{ij}g_{1jk}=\delta^i_{\ k} 
+\epsilon\left(h_{1}-h_{2}\right)^{i}_{\ k}+\epsilon^2\left(h_2^{2}-h_2 h_1\right)^i_{\ k}
+ \epsilon^3\left(h_{2}^2h_1-h_{2}^3\right)^i_{\ k},
\end{equation}
so that
\begin{equation}
\begin{aligned}
\XX^i_{\ k} &= \left(\sqrt{g_2^{-1}g_1}\right)^i_{\ k }=\delta^{i}_{\ k}
+\frac{1}{2}\epsilon (h_{1} - h_{2})^i_{\ k} + \frac{1}{8}\epsilon^2 
\left[ 3 h_{2}^2 - h_{1}^2+  h_1 h_{2} - 3h_2 h_{1} \right]^i_{\ k} \\
&
+ \frac{1}{16}\epsilon^3\left[h_1^3 + h_2 h_1^2 -h_1^2 h_2 -h_2 h_1 h_2 +  h_ 1h_2h_1 + 5 h_2^2 h_1 
-h_1 h_2^2 -5 h_2^3 \right]^i_{\ k}.
\end{aligned}
\end{equation}
Finally, we can expand all traces, 
\begin{equation}
\begin{aligned}
\mathrm{Tr}(\mathbb{X})&=4+\frac{1}{2}\epsilon\mathrm{Tr}(h_1-h_2)+\frac{1}{8}\epsilon^2 \left[3\mathrm{Tr}(h_2^2)-\mathrm{Tr}(h_1^2)-2\mathrm{Tr}(h_1h_2)\right]\\
	&
	+\frac{\epsilon^3}{16}
	\left[
	 \mathrm{Tr}(h_1^3)
	+\mathrm{Tr}(h_2h_1^2)
	+3\mathrm{Tr}(h_2^2h_1)
	-5\mathrm{Tr}(h_2^3)\right] = 	e_1(\mathbb{X}), \\
\mathrm{Tr}(\mathbb{X}^2) &=
4+
\epsilon \mathrm{Tr}(h_1-h_2)
+\epsilon^2\left[ \mathrm{Tr}(h_2^2) - \mathrm{Tr}(h_1h_2)\right] 
+ \epsilon^3 \left[
\mathrm{Tr}(h_2^2h_1)-\mathrm{Tr}(h_2^3)
\right], \\
	\mathrm{Tr}(\mathbb{X}^3)&=
	4+
	\frac{3}{2}\epsilon\mathrm{Tr}(h_1-h_2)
	+\frac{3}{8}\epsilon^2 \left[ 	
	5\mathrm{Tr}(h_2^2)
	+ \mathrm{Tr}(h_1^2)
	- 6\mathrm{Tr}(h_1h_2)	\right]\\
	&
+\frac{1}{16}\epsilon^3
\left[
- \mathrm{Tr}(h_1^3)
-35 \mathrm{Tr}(h_2^3)
-9 \mathrm{Tr}(h_1^2h_2)
+45 \mathrm{Tr}(h_1h_2^2)
\right],
\end{aligned}
\end{equation}
and in the end we have the expansion of all invariants $e_k$:
\begin{equation}
	\begin{aligned}
		e_2(\mathbb{X})&=6 + \frac{3}{2} \epsilon \,  \mathrm{Tr}(h_1-h_2)
		+ \frac{1}{8} \epsilon^2 \, \left[ 	(\mathrm{Tr}(h_1))^2 
		+ (\mathrm{Tr}(h_2))^2 - 2 \mathrm{Tr}(h_1)\mathrm{Tr}(h_2) \right. \\
		& \left. \qquad  \qquad
		+ 8 \mathrm{Tr}(h_2^2) - 4 \mathrm{Tr}(h_1^2) - 4 \mathrm{Tr}(h_1h_2) \right] \\
		& + \frac{1}{16}\epsilon^3\, 
		\left[ 
		4 \mathrm{Tr}(h_1^3)+
		4 \mathrm{Tr}(h_2^2 h_1)
	   -12 \mathrm{Tr}(h_2^3)
		+4 \mathrm{Tr}(h_2 h_1^2)
		 \right.  \\
		&\left. \qquad \qquad 
		 - \mathrm{Tr}(h_1) \mathrm{Tr}(h_1^2)
		  -2 \mathrm{Tr}(h_1) \mathrm{Tr}(h_1 h_2) 
		  + 3 \mathrm{Tr}(h_1) \mathrm{Tr}(h_2^2) \right. \\
		 	&\left. \qquad \qquad 
		 + \mathrm{Tr}(h_2) \mathrm{Tr}(h_1^2)
		 +2 \mathrm{Tr}(h_2) \mathrm{Tr}(h_1 h_2) 
		 - 3 \mathrm{Tr}(h_2) \mathrm{Tr}(h_2^2)
		 \right],
	\end{aligned}
\end{equation}

\begin{equation}
	\begin{aligned}
		e_3(\mathbb{X})&= 
		4  + \frac{3}{2}\epsilon\, \mathrm{Tr}(h_1-h_2) 
		    + \frac{1}{8}\epsilon^2 \, \left[
	- 5 \mathrm{Tr}(h_1^2) 
	- 2	\mathrm{Tr}(h_1 h_2)
 + 7 \mathrm{Tr}(h_2^2) \right. \\
& \left. \qquad \qquad + 2 (\mathrm{Tr}(h_1))^2 + 2 (\mathrm{Tr}(h_2))^2 
 - 4 \mathrm{Tr}(h_2) \mathrm{Tr}(h_1)  \right] \\
&+ \frac{1}{48}\epsilon^3 \, \left[ 
   17  \mathrm{Tr}(h_1^3) 
- 29 \mathrm{Tr}(h_2^3)
+ 9 \mathrm{Tr}(h_1^2h_2)
+ 3 \mathrm{Tr}(h_1h_2^2) 
\right.  \\ & \left. \qquad \quad 
- 9 \mathrm{Tr}(h_1)\mathrm{Tr}(h_1^2)
- 6 \mathrm{Tr}(h_1)\mathrm{Tr}(h_1 h_2)
+ 15 \mathrm{Tr}(h_1)\mathrm{Tr}(h_2^2)
+ (\mathrm{Tr}(h_1))^3
\right.  \\ & \left. \qquad \quad 
+ 9 \mathrm{Tr}(h_2)\mathrm{Tr}(h_1^2)
+ 6 \mathrm{Tr}(h_2)\mathrm{Tr}(h_1 h_2)
- 15 \mathrm{Tr}(h_2)\mathrm{Tr}(h_2^2)
\right.  \\ & \left. \qquad \quad 
- 3 \mathrm{Tr}(h_2) (\mathrm{Tr}(h_1))^2 
+ 3 \mathrm{Tr}(h_1) (\mathrm{Tr}(h_2))^2 
- (\mathrm{Tr}(h_2))^3  \right].
\end{aligned}
\end{equation}
For the last ivariant, arising from the determinant, we have,
\begin{equation}
\begin{aligned}	
 \mathrm{Det}(1+\epsilon A + \epsilon^2 B +\epsilon^3 C) =&
1 + \epsilon \mathrm{Tr}(A) 
+ \epsilon^2 \left( \mathrm{Tr}(B) + \frac{1}{2} \left( (\mathrm{Tr}(A))^2 - \mathrm{Tr}(A^2) \right) \right) \\
& + \epsilon^3 \biggl(
\mathrm{Tr}(C) + 
\mathrm{Tr}(A) \mathrm{Tr}(B) - \mathrm{Tr}(AB)\\
&  \quad 
+\frac{1}{6}( \mathrm{Tr}(A))^3 -  \frac{1}{2} \mathrm{Tr}(A^2) \mathrm{Tr}(A) 
+\frac{1}{3} \mathrm{Tr}(A^3) \biggr),
\end{aligned}
\end{equation}
leading to,
\begin{equation}
\begin{aligned}
e_4(\mathbb{X})&=
1+\frac{1}{2}\epsilon \mathrm{Tr}(h_1-h_2) +\frac{1}{8}\epsilon^2\left(\left(\mathrm{Tr}(h_1-h_2)\right)^2
  + 2\mathrm{Tr}\left(h_2^2-h_1^2\right)\right)\\
  & +\frac{1}{48}\epsilon^3\left[8\mathrm{Tr}(h_1^3-h_2^3)+\mathrm{Tr}(h_1-h_2)\left(6\mathrm{Tr}(h_2^2-h_1^2)+\left(\mathrm{Tr}(h_1-h_2)\right)^2\right)\right].
\end{aligned} 
\end{equation}

As a result the $\epsilon^2$-part of the interaction in the above action is of the form (we omit higher order 
corrections here, as the formula gets too complicated and not transparent):

\begin{equation}
\begin{aligned}
S_{\mathrm{int}}^{(2)} 
& \sim (\beta_0 + 4 \beta_1 + 6 \beta_2 + 4 \beta_3 + \beta_4)  \\
& + \frac{1}{2} \epsilon (\beta_1 + 3\beta_2 + 3\beta_3 + \beta_4) \mathrm{Tr}(h_1) \\
& + \frac{1}{2} \epsilon (\beta_0 + 3\beta_1 + 3\beta_2 + \beta_3) \mathrm{Tr}(h_2) \\
& - \frac{1}{8} \epsilon^2 \left( \beta_3 +4 \beta_2 + 5\beta_1 +2 \beta_0\right) \mathrm{Tr}(h_2^2) \\
& - \frac{1}{8} \epsilon^2 \left( \beta_1 +4 \beta_2 + 5 \beta_3 +2 \beta_4\right) \mathrm{Tr}(h_1^2) \\
& - \frac{1}{8} \epsilon^2 \left( 2\beta_1 +4 \beta_2 +2 \beta_3 \right) \mathrm{Tr}(h_1 h_2) \\
& + \frac{1}{8} \epsilon^2 \left( \beta_2 + 2\beta_3 + \beta_4 \right) \left(\mathrm{Tr}(h_1)\right)^2 \\
& + \frac{1}{8} \epsilon^2 \left( \beta_0 + 2\beta_1 + \beta_2 \right) \left(\mathrm{Tr}(h_1)\right)^2 \\
& + \frac{1}{4} \epsilon^2 \left( \beta_1 + 2\beta_2 + \beta_3 \right) \left(\mathrm{Tr}(h_1)\mathrm{Tr}(h_1)\right) \\
& + \epsilon^3 \cdots
\end{aligned} 
\label{betaexp}
\end{equation}

Comparing the above expression with Sec. \ref{Sh1h2} we see that there are two quadratic terms 
in  Sec. \ref{betaexp} that have the same coefficient proportional to $ \beta_1 + 2\beta_2 + \beta_3$ but one of them vanishes  in Sec. \ref{Sh1h2} and the other does not. As a conclusion, even the perturbative form of the spectral action for the interacting geometries cannot be equivalent
to the usually assumed model of action for bimetric gravity.
 
Similarly, one can demonstrate that a second natural choice to identify $h_1$ and $h_2$ from the bimetric perturbative
expansion with  $W_-$ and  $W_+$  fields also leads to contradiction already in the second order of the expansion
in $\epsilon$.

\section{The interaction in terms of invariants}
Even though the usually assumed form of the action for the bimetric gravity is not compatible with the spectral
interactions between geometries one has to observe that the assumed form of the action is very restrictive as
it uses only the coefficients of the invariant polynomial of the matrix $\XX$. This particular choice is quite elegant,
yet it restricts a lot the possible interaction terms. 

A natural question is, what is the invariant form of the interaction in the perturbative expansion, which is expressed
as polynomials in the invariants of the matrix $\XX$. As there are only four independent invariants, we assume that
the perturbative action is polynomial of order at most $4$ in $\XX$:
\begin{equation}
\label{termsINV}
\begin{aligned}
S_{\mathrm{int}}\sim \int d^4x & \sqrt{\det g_2} \left(\alpha_0
+ \alpha_1 \mathrm{Tr'}(\mathbb{X})
+\alpha_2 \mathrm{Tr'}(\mathbb{X}^2)
+\alpha_3\left(\mathrm{Tr'}(\mathbb{X})\right)^2 
+\alpha_4 \mathrm{Tr'}(\mathbb{X}^3)\right.\\ &\left. 
+\alpha_5 \mathrm{Tr'}(\mathbb{X})\mathrm{Tr'}(\mathbb{X}^2)
+\alpha_6\left(\mathrm{Tr'}(\mathbb{X})\right)^3
+\alpha_7\mathrm{Tr'}(\mathbb{X}^4)
+\alpha_8 \mathrm{Tr'}(\mathbb{X}) \mathrm{Tr'}(\mathbb{X}^3)\right. \\
&\left. +\alpha_{9}\left(\mathrm{Tr'}(\mathbb{X}^2)\right)^2
+\alpha_{10}\mathrm{Tr'}(\mathbb{X}^2)\left(\mathrm{Tr'}(\mathbb{X})\right)^2
+\alpha_{11}\left(\mathrm{Tr'}(\mathbb{X})\right)^4\right),
\end{aligned}
\end{equation}
where for convenience we use $\hbox{Tr}' = \hbox{Tr} - 4$. 
The only term, which we did not expand earlier is the last one, $\mathrm{Tr}(\mathbb{X}^4)$, and
its expansion up to third order in $\epsilon$, is
\begin{equation}
\begin{aligned}
\mathrm{Tr}(\mathbb{X}^4)&=4+2\epsilon\mathrm{Tr}(h_1-h_2)+\epsilon^2\left(\mathrm{Tr}(h_1^2)+3\mathrm{Tr}(h_2^2)-4\mathrm{Tr}(h_1h_2)\right)\\
&+\epsilon^3\left(-4\mathrm{Tr}(h_2^3)+6\mathrm{Tr}(h_1h_2^2)-2\mathrm{Tr}(h_1^2h_2)\right).
\end{aligned}
\end{equation}
We expand \eqref{termsINV} in $\epsilon$ and compare it (up to order $\epsilon^3$) with \eqref{Sh1h2}. This leads 
to a linear system of equations that has a four-parameter family of solutions, with
\begin{equation}
\begin{aligned}
\alpha_1&=-16+\alpha_4,\qquad & \qquad \alpha_2=10-\frac{3}{2}\alpha_4,\\
  \alpha_3&=-2-\alpha_5,\qquad & \qquad \alpha_7=-1-\frac{1}{4}\alpha_4,\\
  \alpha_9&=\frac{1}{2}-\frac{1}{4}\alpha_5-\frac{3}{4}\alpha_8,\qquad & \qquad 
\alpha_{10}=-\frac{1}{2}\alpha_6,
\end{aligned}
\end{equation}
and $\alpha_0=0$.

It is, in particular, possible to find a unique solution in the form of a polynomial of the lowest order in $\XX$, which
is a polynomial of third order, an the resulting action reads,
\begin{equation}
S_{\mathrm{int}}\sim \int d^4x \sqrt{\det g_2} 
\biggl(-10 \mathrm{Tr'}(\mathbb{X}) +8 \mathrm{Tr'}(\mathbb{X}^2) 
-2\left(\mathrm{Tr'}(\mathbb{X})\right)^2 -2 \mathrm{Tr'}(\mathbb{X}^3)+ \mathrm{Tr'}(\mathbb{X})\mathrm{Tr'}(\mathbb{X}^2)\biggr).
\end{equation}
In other words, in the third order in $\epsilon$ we can eliminate all the terms of order higher than three in $\mathbb{X}$.
Passing back to traces the formula is even simpler,
\begin{equation}
	S_{\mathrm{int}}\sim \int d^4x  \sqrt{\det g_2} 
		\biggl(2 \mathrm{Tr}(\mathbb{X}) +4 \mathrm{Tr}(\mathbb{X}^2) 
		-2\left(\mathrm{Tr}(\mathbb{X})\right)^2  -2 \mathrm{Tr}(\mathbb{X}^3)+ \mathrm{Tr}(\mathbb{X})\mathrm{Tr}(\mathbb{X}^2)\biggr).
\end{equation}
\section{Conclusions and outlook}

Having derived an explicit perturbative form of the {\em interaction} term between geometries using the spectral methods for a simple two-sheeted 
non-product geometry we find that although it resembles the bimetric
gravity theory the coefficients of the interaction potential cannot be 
matched to such a model. 

There are, however, many interesting features of our result. First of all,
it confirms (perturbatively up to the third-order) that the nonlinear interaction term between the geometries is expressed through a function of the invariants of the matrix $\mathbb{X} =\sqrt{g_2^{-1} g_1}$. Although this is almost obvious, due to the general covariance of the spectral action, no explicit formula for this function is known. Here, we find its perturbative expansion around flat geometry, which can be used to study the stability of interacting geometries and cosmology models. This formulation opens also the possibility for further examination of the ghost problem in our model. At first, comparing the perturbative form of the action to Fierz-Pauli  theory \cite{FP} one guesses that there will be ghosts as the action has only one quadratic term. Yet, the full analysis is more intricate and we postpone the detailed studies for the future.  
 	
Furthermore, the explicit form of perturbative terms is quadratic in the difference of the small perturbations, which indicates that the flat geometry is indeed stable. Moreover, only one of the linearized fields will be massive and interact with the massless linear perturbations. 

The main result of the paper is that there exists a natural, canonical and {\em geometric} interaction between two adjacent geometries. Independently of the interpretation that relates it to brane interactions in the bulk, interacting universes, bimetric gravity or noncommutative geometry, the interaction is fixed in the same way the invariance fixes the usual action terms for gravity (the cosmological constant and the Einstein-Hilbert scalar curvature term).  It is an open intriguing question of what are the physical consequences of such interactions between geometries and what effects they have on cosmology. 

\appendix
\section{Polynomial integrals over higher spheres}
\label{app:A}
We review here technical tool of the computation, which are the integrals
of polynomial functions over the unit spheres. We are interested in the value 
of the following quantity, 
\begin{equation}
I_{n,m}^{\alpha_1\beta_1...\alpha_m\beta_m}=\int_{\|x\|=1}d^nx \, x^{\alpha_1}x^{\beta_1}...x^{\alpha_m}x^{\beta_m},
\end{equation}
i.e. the monomial integrals over a unit sphere. This can be done by the straightforward generalization into higher dimensional cases of the method presented in \cite{Ot11} for the $2$-sphere (see also \cite{Ba97},\cite{Fo01}).
By denoting 
\begin{equation}
\gamma_j=\begin{cases}\alpha_k, \qquad j=2k-1 \\ \beta_k, \qquad j=2k\end{cases} 
\end{equation}
for $k=1,...,m$, we then have 
\begin{equation}
I_{n,m}^{\alpha_1\beta_1...\alpha_m\beta_m}\equiv I_{S_n}^{\gamma_1...\gamma_{2m}}=\int_{\|x\|=1}d^nx x^{\gamma_1}...x^{\gamma_{2m}}.
\end{equation}

Let $S_n=\partial B_n\equiv S^{n-1}$ in $\mathbb{R}^n$. The following generalization of \cite[Prop.~1]{Ot11}, which can be proven by induction on $m$, holds:
\begin{proposition}
Let $I_{B_n}^{\gamma_1...\gamma_{2m}}=\int_{\|x\|\leq 1}d^nx x^{\gamma_1}...x^{\gamma_{2m}}$. Then
\begin{equation}
I_{B_n}^{\gamma_1...\gamma_{2m}}=\frac{1}{2m+n}I_{S_n}^{\gamma_1...\gamma_{2m}}.
\end{equation}
\end{proposition}
Similarly, \cite[Prop.~2]{Ot11} can be easily generalized to arbitrary dimensions:
\begin{proposition}
\label{prop_m}
\begin{equation}
I_{S_n}^{\gamma_1...\gamma_{2m+2}}
=\frac{1}{2m+n}\left[\delta^{\gamma_1\gamma_2}
I_{S_n}^{\gamma_3...\gamma_{2m+2}}
+...
+\delta^{\gamma_1\gamma_{2m+2}}I_{S_n}^{\gamma_2...\gamma_{2m+1}}\right].
\end{equation}
\end{proposition}
The proof is again based on the induction.

The explicit formulae used in this paper concern three values in the
four-dimesional case, which we present explicitly,
\begin{equation}
I^{\gamma_1...\gamma_{2m}}\equiv I_{S_4}^{\gamma_1...\gamma_{2m}}=\int_{\|x\|= 1}d^4x x^{\gamma_1}...x^{\gamma_{2m}}.
\end{equation}
For $m=0$ we have $I^0=\mathrm{area}(S_4)=2\pi^2$. Now, using Prop. \ref{prop_m}, we immediately get
\begin{equation}
I^{\gamma_1\gamma_2}=\frac{1}{4}\delta^{\gamma_1\gamma_2}\mathrm{area}(S_4)=\frac{\pi^2}{2}\delta^{\gamma_1\gamma_2}
\end{equation}
\begin{equation}
I^{\gamma_1\gamma_2\gamma_3\gamma_4}=\frac{\pi^2}{12}\left[\delta^{\gamma_1\gamma_2}\delta^{\gamma_3\gamma_4}+ \delta^{\gamma_1\gamma_3}\delta^{\gamma_2\gamma_4} + \delta^{\gamma_1\gamma_4}\delta^{\gamma_2\gamma_3}\right],
\end{equation}
and
\begin{equation}
\begin{aligned}
I^{\gamma_1\gamma_2\gamma_3\gamma_4\gamma_5\gamma_6}=\frac{\pi^2}{96}\left[\right.&\left.\delta^{\gamma_1\gamma_2}\left(\delta^{\gamma_3\gamma_4}\delta^{ \gamma_5\gamma_6}+\delta^{\gamma_3\gamma_5}\delta^{ \gamma_4\gamma_6}+\delta^{\gamma_3\gamma_6}\delta^{ \gamma_4\gamma_5}\right)+\right.\\ +& \left. \delta^{\gamma_1\gamma_3}\left(\delta^{\gamma_2\gamma_4}\delta^{ \gamma_5\gamma_6} + \delta^{\gamma_2\gamma_5}\delta^{ \gamma_4\gamma_6}+\delta^{\gamma_2\gamma_6}\delta^{ \gamma_4\gamma_5}\right)+ \right.\\ +&\left.\delta^{\gamma_1\gamma_4}\left(\delta^{\gamma_2\gamma_3}\delta^{ \gamma_5\gamma_6}+ \delta^{\gamma_2\gamma_5}\delta^{ \gamma_3\gamma_6} + \delta^{\gamma_2\gamma_6}\delta^{ \gamma_3\gamma_5}\right)+\right. \\+& \left.
\delta^{\gamma_1\gamma_5}\left(\delta^{\gamma_2\gamma_3}\delta^{ \gamma_4\gamma_6}+ \delta^{\gamma_2\gamma_4}\delta^{ \gamma_3\gamma_6} + \delta^{\gamma_2\gamma_6}\delta^{ \gamma_3\gamma_4}\right)+ \right. \\
+&\left. \delta^{\gamma_1\gamma_6}\left(\delta^{\gamma_2\gamma_3}\delta^{ \gamma_4\gamma_5}+ \delta^{\gamma_2\gamma_4}\delta^{ \gamma_3\gamma_5} + \delta^{\gamma_2\gamma_5}\delta^{ \gamma_3\gamma_4}\right)
 \right].
\end{aligned}
\end{equation}

\noindent{\bf Acknowledgments: \\}
\noindent
A.B. acknowledges the support from the National Science Centre, Poland, Grant No. 2018/31/N/ST2/00701. 
A.S. acknowledges the support from the  National Science Centre, Poland, Grant No. 2020/37/B/ST1/01540.


\end{document}